\documentclass[aps,prl,twocolumn,showpacs,superscriptaddress]{revtex4}
\usepackage{graphics}
\usepackage{epsfig}

\begin{document}

\title{Information Horizons in Networks}

\author{A. Trusina}\email{trusina@tp.umu.se}
\author{M. Rosvall}

\affiliation{Department of Theoretical Physics, Ume{\aa} University,
901 87 Ume{\aa}, Sweden}
\affiliation{NORDITA, Blegdamsvej 17, Dk 2100, Copenhagen, Denmark}\homepage{www.nordita.dk/research/complex}
\author{K. Sneppen}
\affiliation{NORDITA, Blegdamsvej 17, Dk 2100, Copenhagen, Denmark}
\homepage{www.nordita.dk/research/complex}

\date{\today}

\begin{abstract}
We investigate and quantify the interplay between topology and ability to send
specific signals in complex networks.
We find that in a majority of investigated real-world networks
the ability to communicate is 
favored by the network topology on small
distances, but disfavored at larger distances. We further discuss how
the ability to locate specific nodes can be improved if
information associated to the overall traffic in the network is available.
\end{abstract}
\maketitle

Not all different parts interact directly with all
other parts in a complex system.
Rather each
element interacts directly only with a few particular elements.
Distant parts of the thereby formed
network can consequently communicate through sequences
of local interactions. In this way all parts of the network in
principle can be reached from other parts, but not all such
communications are equally easy or accurate.
The network is thus a description of
the limited ability to send specific signals in the
system \cite{rosvall}.
We stress the difference between specific signaling in networks and the contrary unspecific broadcasting:
Where specific signaling only focuses on locating one specific node without
disturbing the remaining network, the
non-specific broadcasting amplifies
by transfering signals to all exit links of every
node along all branching paths.
Specific signaling is thus constructive communication, whereas
non-specific broadcasting rather is of relevance for disease spreading
or computer virus propagation \cite{moreno,newmanVirus}. 

One can imagine various ways of searching 
a specific node in a network,
dependent on the available information
when the search is performed \cite{adamic}. In
present paper we compare ways to guide the search based on locating
the shortest path between a source and a target in the network. Thus
we are only characterizing specific signaling, where any deviations
from shortest paths mean the loss of the signal. In other words, the
cost of deviating from the shortest path is assumed to be infinite,
and we simply quantify the search in terms of the number of
questions needed to follow the shortest path to the target.
\begin{figure}[htbp]
\centerline{\psfig{file=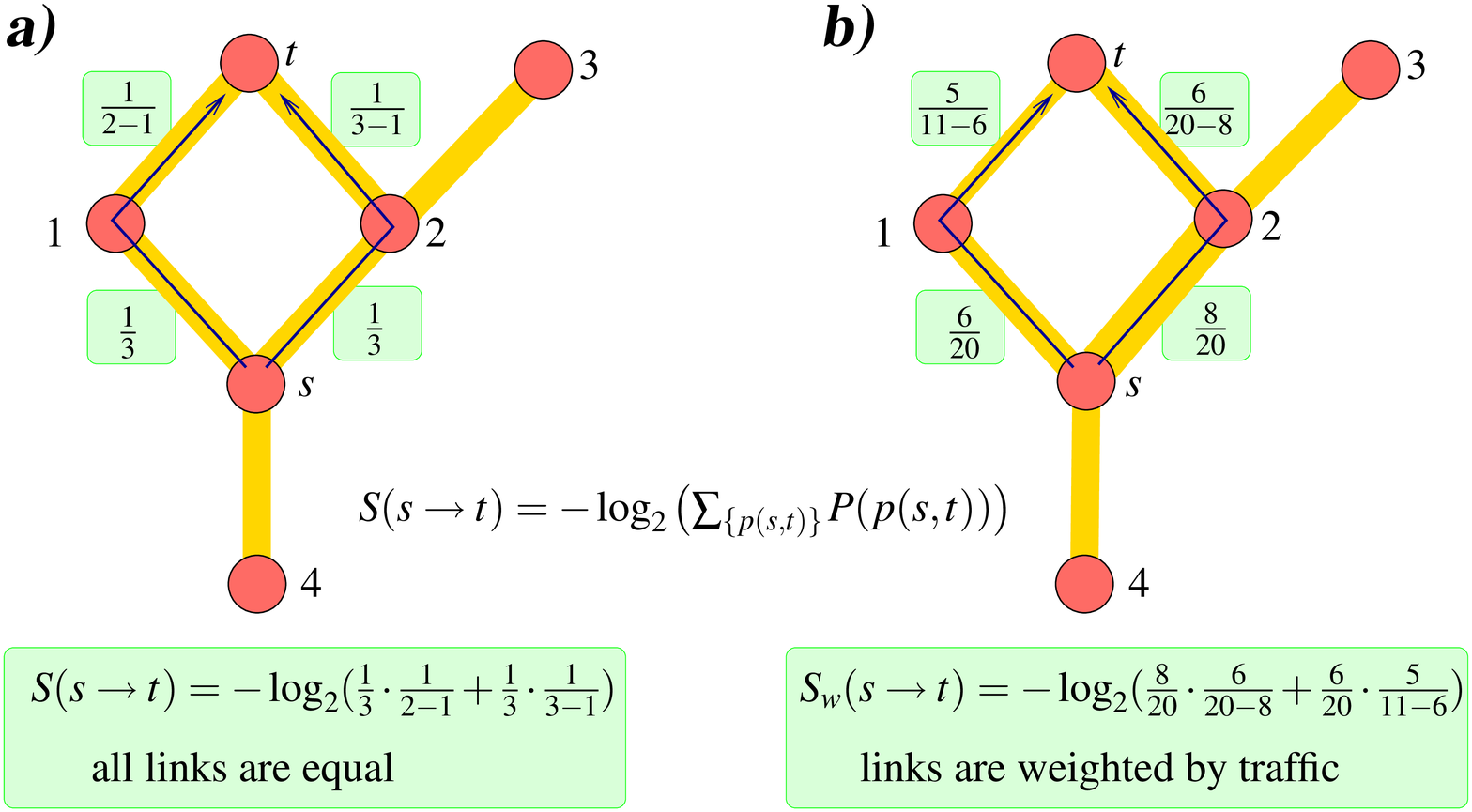,angle=0,width=9cm}}
\caption[] {Information measures on network topology:
{\sl \textbf{a)}} {\it Search Information}
\mbox{$S(s \to t)$} measures your ability to locate node $t$ from
node $s$. {\sl \textbf{b)}} {\it Weighted Search Information} $S_w$
measures your ability to locate target node $t$ from the source node $s$,
when you tend to follow the traffic given by the betweenness $b_{ij}$.
\mbox{$S(s \to t)$} is the number of yes/no questions needed to
locate any of the shortest paths between node $s$ and node $t$. For
each such path $P(p(s,t)) \; =\; \frac{1}{k_s} \;\; \prod_{j}
\frac{1}{k_j-1}$, with $j$ counting nodes on the path $p(s,t)$ until
the last node before $t$. The factor $k_j-1$ instead of $k_j$ takes
into account the information gained by following the path.
$S_w(i \to j)$ is the similar quantity where we now weight each exit
link from a node with its betweenness $\beta_{lk}$
\protect{\cite{freeman,between}}, defined as the fraction of messages that
go through node $l$ which also go through neighbor node $k$. }
\label{fig1}
\end{figure}

First let us consider the Search Information introduced in
\cite{pnas}. The Search Information of going from source node $s$ to
target node $t$, $S(s\rightarrow t)$, is the number of
bits of information one needs to go from $s$ to $t$ using the shortest
paths: In the beginning, when starting at node $s$, one has to find
the right exit link,$leading to$ 
the second node on the shortest path to the target 
node $t$. We assume that each node is a simplistic autonomous system
that knows which of its exit links that leads to
the target. The number of questions one has to ask such an
autonomous system in a source node is $\log_2 (k_s) $, where $k_s$
is the degree of the source. At the subsequent  node, $j$, along the
shortest path to the target the number of questions is reduced to
$\log_2 (k_j-1)$ since the incoming link is known. That means
that the number of questions one has to ask when walking along the
path from the source to the target is $S(s\rightarrow t) =
-\log_2(\frac{1}{k_s}\prod_{j}\frac{1}{k_j - 1})$. If there are more
than one shortest path between $s$ and $t$, then:
\begin{equation}
S(s\rightarrow t) =
-\log_2\left(\sum_{\{p(s,t)\}}\frac{1}{k_s}\prod_{j}\frac{1}{k_j - 1}\right),
\end{equation}
where the sum runs over the set $\{p(s,t)\}$ of degenerate shortest paths between
$s$ and $t$, see Fig.\ \ref{fig1}.
\begin{figure}[htbp]
\centerline{\psfig{file=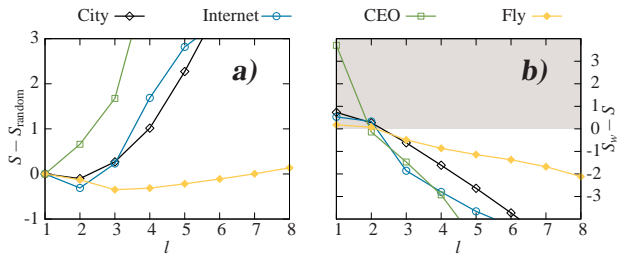,angle=0,width=9cm}}
\caption[] {Analysis of real world networks. Here City is 
the information city network of Malm\"o \protect{\cite{city}} with
roads mapped to nodes and intersections to links. The Internet refers
to the hardwired networks of autonomous systems
\protect{\cite{internet}}, the CEO to the network of cooperate
executives in US \protect{\cite{ceo}} and Fly to the protein-protein
network of Drosophilia Melanogaster \protect{\cite{giot}}. In {\sl \textbf{a)}} we
compare $S(l)$ with the similar search performed in a randomized
version of the network. One observes that search on short distances
$l\sim 2-3$ is relatively optimized in the real networks. In {\sl \textbf{b)}} we
compare $S$ with the search obtained when one uses the information
associated to overall traffic in the network. We see that such
global traffic information helps the search at all long distances. }
\label{fig2}
\end{figure}
In the previous work \cite{pnas} we investigated $S$ for a number of
networks and found that one needs more information to orient 
in real- than in random networks. By random networks we mean the networks
randomized by the reshuffling of links in such a way as to preserve
the degree sequence and keep the network connected
\cite{maslov2002}.
To explore the nature of these complications in real world networks we, in
Fig.\ \ref{fig2}, look at the average Search Information for nodes
separated by $l$ links and compare it with the corresponding
quantity in a randomized counterpart. From  $\Delta S(l) = \langle
S(l) \rangle - \langle S_{random}(l) \rangle$ we see that
essentially all the contribution to the global excess of $\Delta S =
S - S_{random}$ comes from large distances $l > 3$ ($S=\frac{1}{N^2}\sum_{s,t}S(s \to t))$. For some of the
networks, as for example Internet, Yeast and Fly, the $\Delta S(l)$ is
even negative at short distances, which implies that these real
networks are organized to optimize the search at
these short distances. Thus local specificity is favored whereas
communication beyond the horizon $l=l_{search}\sim 3$ is disfavored.

\begin{figure}[htbp]
\centerline{\psfig{file=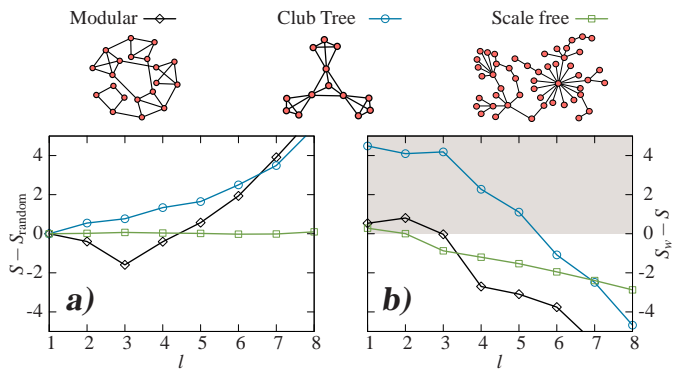,angle=0,width=9cm}}
\caption[] {Analysis of model networks in terms of the
quantities in Fig.\ \ref{fig2}. The simple modular network is
constructed of $C$ modules with $C$ nodes in each, with $0.2C$
connections between nodes in modules and $0.2C$ connections between
the modules ($C=\sqrt{N}$). The club tree is a hierarchical network
with club structure at each level, a construction intended to mimic
our view of social organization. We used a version with $\langle k \rangle =6$
neighbors per node and with $\approx log(N)/log(\langle k \rangle/2)$ hierarchical
levels. The scale-free network is an example of networks with
broad degree-distributions, here scaling as $1/k^{2.4}$. In all cases
we simulate $N=5000$ node networks. } \label{fig3}
\end{figure}

To uncover how the topology and the search information, quantified by $S$, are coupled
we, in Fig.\ \ref{fig3} investigate a number of model networks. Fig.\ \ref{fig3}(a) 
shows $S(l)-S_{random}(l)$ for the various model
networks. We see that the search is easy at  distance $l\sim 3$ in
the modular network, whereas
a randomly rewired network provides better search options for $l>3$.
The hierarchical club network, on the other hand, clearly does worse
than a random network on all scales. Here we have obtained a
surprising and counterintuitive result that hierarchies are not
always optimal for search.
That (club)hierarchies are used in many
human organizations may thus be seen as a way to regulate
and thus limit the information exchange, rather than
to optimize overall specific communication \cite{footnote}.


The search information $S$ defined above is based on a minimal
approach where one at each node knows nothing about the relative
importance of the neighbors. However, in real social networks one often knows
who is best connected to the rest of the system.
For example in a military hierarchy, 
every soldier knows who their immediate superior
is. This knowledge can be obtained self consistently at any node in
any network by monitoring the traffic of orders past this node.
In order to explore how the search can be simplified by additional
knowledge we introduce a
slightly different quantification of search information.
That is, we explore the information
needed to search if one knows the overall traffic flow.
When questioning
the minimal autonomous system at a node, we weight the questions
according to the betweenness of the links to the node \cite{freeman,between}. Thereby we
define the weighted search information
\begin{equation}
S_w(s\rightarrow t) = - \log_2 \left( \sum_{\{p(s,t)\}} b_{s,j=1}
\prod_{j\in p(s,t)} b'_{j,j+1} \right), \label{weight}
\end{equation}
where $j$ labels the node on the path $p(s,t)$, starting at
$j=1$ for neighbor node to $s$. $b_{j,j+1}=\beta_{j,j+1}/\sum_k
\beta_{j,k}$ is the betweenness of the link from node with label
$j$ to node with label $j+1$, divided by the sum of the betweennesses of all  $k$ 
links from $j$. $b'_{j,j+1}$ is similarly defined except that
the normalization excludes the link to the preceding node of $j$ on the shortest path between $s$ and $t$.

To understand the difference between $S$ and $S_w$ we consider a
city (defined through the city network where each node is a road,
and each link an intersection \cite{city}).
By orienting yourself with the strategy behind $S$ small and
large roads are weighted equally. However, $S_w$ captures that large
roads more often take you closer to the target than small roads. For
all investigated networks one in average gains by using the weighted search
strategy. 
However, the contribution is not
homogeneously distributed over distance. As one can see in Fig.\ \ref{fig2}(b)
the weighted strategy is more efficient at longer
distances, $l> 3$. However $S_w >S $ for $l \leq 3$ and thus it
turns out to be inefficient to follow the flow when the target is
nearby. 
This reflects the fact that if you follow the flow you
will nearly always overlook small roads in your neighborhood. In
terms of navigating in a city, the $S_w(l)-S(l)$ difference shows
that it pays off to follow the large roads until you are within a
few turns from your end target. Then it
naturally pays off to change strategy and disregard the main stream.
The distance where $S_w(l)-S(l)$ becomes negative therefore defines
a characteristic search horizon, $l_{local-global}$, at which one
should switch from local to global search strategy.

We next study the relative advantage of local
versus global search strategies for some model
networks in Fig.\ \ref{fig3}(b). 
Like the real world
networks, also the model networks have $S_w>S$ at small distances,
and $S_w<S$ at large distances. In particular, the club tree (hierarchy)
does extremely bad at short distances because 
there is a strong bias to go
along the main flow, and one thus needs a lot of effort to locate
peripheral neighbors. For a random scale-free network, on the other
hand, the overall traffic very fast guides you to the center, and
therefore $S_w$ is a good search strategy at nearly all distances.
The scale-free network
represents topologies with very broad degree distributions and in
these one nearly always benefit by following the flow \cite{adamic}.

In between
these two networks is the modular network, where the global flow
confuses local search ($S(l<3) < S_w(l<3)$), but helps traffic to
other modules and thus to the more distant targets.
Returning to the real world
networks in Fig.\ \ref{fig2}(b) their $l_{local-global}\sim 2$ horizon
for traffic guided search may be seen as a combination of a short
$l_{local-global}\sim 1$ horizon associated to their broad degree
distribution (scale free in Fig.\ \ref{fig3}(b)), and a larger
$l_{local-global}$ horizon associated to modular or hierarchical
features.

\begin{figure}[htbp]
\centerline{\psfig{file=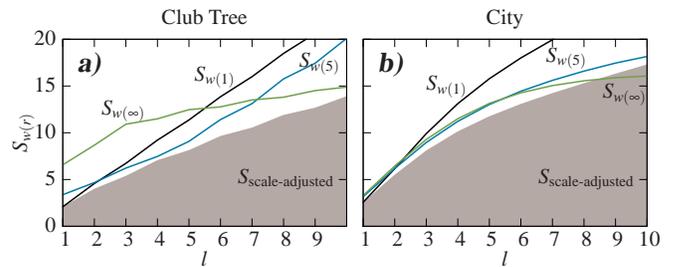,angle=0,width=9cm}}
\caption[] {Investigations of search strategies in a model network {\sl \textbf{a)}} and
information city network of Malm\"o {\sl \textbf{b)}}.
The figures illustrate 3 simple search strategies, and one
optimized (shaded). $S_{w(r)}(l)$ denotes the search information
based only on traffic between nodes separated by not more
than $r$ steps in the network.
We see that nodes on short distance are best found by using
local search strategy ($S_{w(r=1)}=S$, whereas
search to distant nodes are best performed by using
information  from global traffic.
However, nodes at intermediate distances are best found
by using traffic between nodes at intermediate distances.
To optimize search we also show a different search strategy $S_{\mathrm{scale-adjusted}}$, where
one at each step $j$ along the path
to the target adjust the traffic horizon
to the remaining distance to the target.
We furthermore restrict the traffic bias to
the subset of traffic that are targeted to
the node $j$ one is currently at (see Fig.\ \ref{fig5}).
}
\label{fig4}
\end{figure}

One may ask whether the two search strategies can be combined, such
that one uses local information for local search, and global traffic
information for long distance search. In terms of traffic in a city
the picture is that there are multiple types of traffic, from
pedestrian to short distance targets, bicycles to intermediate distance targets, to cars for the distant targets. 
In accordance to this picture we
introduce the limited betweenness measures $b_{ij}(r)$ for the links $j$
around a node $i$, defined by
traffic between all pairs of nodes
that only moves at maximum a distance $r$ between the 
source and the target.
Given this set of $r$ dependent traffic weights, we in
analogy with Eq.\ \ref{weight}, define a set of search measures
$S_{w(r)}(l)$. For $r=1$, $S_{w(1)}(s\rightarrow t)=S(s\rightarrow
t)$, whereas $S_{w(r=\infty)}(s\rightarrow t)=S_w(s\rightarrow t)$
and thus $S_{w(r)}$ naturally interpolates between the non-weighted
and the traffic-weighted search approaches. In Fig.\ \ref{fig4}(a) we
examine $S_{w(r)}(l)$ for the club-hierarchy network from Fig.\ \ref{fig3}. 
In accordance with Fig.\ \ref{fig3}(a) we again
see that the longer distances indeed are best
searched by using long distance traffic.
In addition we see that intermediate distances
$l=3-7$ are best
searched by using a search weighted by traffic traveling
intermediate distances $r\sim 5$, as quantified by $S_{w(5)}(l)$.

\begin{figure}[htbp]
\centerline{\psfig{file=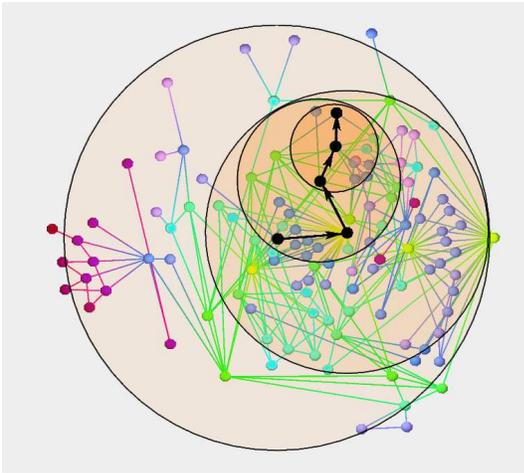,angle=0,width=7cm}}
\caption[] {Illustration of optimal search strategy on an information city
network of Gamla stan, Stockholm \protect{\cite{city}}:
At each step the contribution to the traffic bias is limited
by sequentially decreasing horizons (circles).
The radius of each horizon reflects the node distance to the target.
In case of the large city, each horizon corresponds to the type of
transportation one should consider: Within the big circle one looks for cars
and within the smallest circle for pedestrians.
The path is indicated in black.}
\label{fig5}
\end{figure}

Fig.\ \ref{fig4}(b) shows the optimal
search strategy in a real network, here the information city
network of Malm\"o \cite{city}.
Again the search efficiency is improved
by adjusting the traffic horizon to the search distance.
In fact the search can be further
optimized by, at each step, adjusting the traffic horizon $r$ to the
remaining distance to the target. In the language of city networks,
when searching a distant road, one first uses information from car
traffic, but as distance to target becomes smaller than say 5
intersections, one instead uses bicycle- and then subsequently
pedestrian traffic.
This overall feature of optimizing search works best
when one weights the exit link from each node
$j$ by the fraction of overall traffic target
explicitly to $j$. Thus, the optimal search
indicated by the shaded area in Fig.\ \ref{fig4}
corresponds to a search strategy, where one at each step $j$
from $s$ to $t$ bias the
search according to the subpart of the traffic
that is targeted at $j$, and which has a source
at distance that are not further away than the target $t$
(see Fig.\ \ref{fig5}). The difference to the normal betweenness is
that the target betweenness 
effectively partitions the network
around each node $j$, such that each exit link is weighted
by the fraction of the network that it leads to.
This therefore provides a more efficient
guess on the direction to the rest of the network from $j$
than the normal betweenness.

Obviously the optimal search strategy
can only be used if one has access to this distant-dependent traffic
information. 
However, as in a city, such information can
for example be quite well
estimated in social networks.
Consider Milgram's famous result of
a mail locating a target person in a chain of typically 
six acquaintances between two persons in USA \cite{milgram}.
The nontrivial result of
Milgram's experiment is not that the distance between two persons is just six,
since the dimension in social networks are high \cite{kochen},
but the fact that short paths were found in the experiment.
In terms of our optimal search strategy, Milgram's
experiment is interpreted the following way:
Every participant that receives a mail aimed to
a distant target person, gives this in his turn to a
friend, with a chance weighted to how often this friend travels
on distances up to the scale of the target distance.
With such a search strategy, that at each point along the path
is adjusted to the horizon to the target, the mail will
find a short path to the target person with high probability
(low information cost).
We speculate that humans inherently tend to use such a scale-free
search strategy, and by this facilitate robust communication
on all scales ranging from a single remote village
to the whole planet. The information gain by doing so
in the city Malm\"o is
illustrated by the difference between the black curve and the shaded
area in Fig.\ \ref{fig4}(b).

In the present work we have quantified the information cost
associated to transmission of specific signals across a complex
network. By comparing real- and random networks, we have shown 
that many real-world networks tend to have optimized searchability at
rather short distance $l\sim 3$.
The cost of this optimization is that beyond this horizon one must use 
more intelligent methods to facilitate searchability.
In the spirit of communication we have investigated methods based on global traffic observed at local level and interpreted them in real-world examples.

In many networks, in particular
social or traffic networks, the search strategy can be adjusted
according to average traffic flow. The distance at which global
traffic becomes superior to unbiased search defines a horizon
associated to the largest scale of modules in a network.
In general, any network we have investigated are best searched by
using the ``scale invariant" strategy, where directions are selected
according to the average traffic to nodes at distances
similar to that of the searched target node.

\vfill\eject


\begin{thebibliography}{10}
\vskip -0.5cm

\bibitem{rosvall}
M. Rosvall and K. Sneppen,
Phys. Rev. Lett. {\bf 91}, 178701 (2003).

\bibitem{moreno}
Y. Moreno and A.Vazquez. Eur. Phys. J. B. {\bf 31} 265 (2003).

\bibitem{newmanVirus}
J. Balthrop, S. Forrest, M. E. J. Newman, and M. M. Williamson, Science {\bf 304}, 527 (2004).

\bibitem{adamic}
L. Adamic, B. Huberman, R. Lukose and A. Puniyani,
Phys. Rev. E {\bf 64}, 46135 (2001).

\bibitem{freeman}
L. C. Freeman,
Sociometry {\bf 40}, 35 (1977).

\bibitem{between}
M. E. J. Newman.
Phys. Rev. E {\bf 64}, 016132 (2001).

\bibitem{pnas}
K. Sneppen, A. Trusina and M. Rosvall,
cond-mat/0407055 (2004).

\bibitem{city}
M. Rosvall, A. Trusina and K. Sneppen,
cond-mat/0407054 (2004).

\bibitem{ceo}
G. F. Davis and H.R. Greve,
American Journal of Sociology {\bf 103}, 1 (1997).

\bibitem{giot}
L. Giot et al.,
Science {\bf 302}, 1727 (2003).

\bibitem{maslov2002}
S. Maslov and K. Sneppen,
Science  {\bf 296}, 910 (2002).

\bibitem{milgram}
S. Milgram Psychol. Today 1, {\bf 61} (1967).

\bibitem{kochen}
M. Kochen, Ed., ``The Small World'' (Ablex, Norwood, 1989).

\bibitem{internet}
Website maintained by the NLANR Measurement and
Network Analysis Group at http://moat.nlanr.net/



\bibitem{footnote}
The hierarchical search algorithms are however effective,
but only in the specific case of perfect tree-like hierarchy
searched from the top. In that case the $S$ measured from the top is
$\sim log_2(N+1)-1-2/N+4log_2(N+1)/(N(N+1))\approx log_2(N/2)$ which
is a smaller search information than for any other
organization of a network consisting of $N$ nodes.

\bibitem{acknow}
We acknowledges the support of Swedish Research Council through
Grants No. 621 2003 6290 and 629 2002 6258.

\end{thebibliography}
\end{document}